\documentclass[lettersize,journal]{IEEEtran}
\usepackage{amsmath,amsfonts}
\usepackage{algorithmic}
\usepackage{algorithm}
\usepackage{array}
\usepackage[caption=false,font=normalsize,labelfont=sf,textfont=sf]{subfig}
\usepackage{textcomp}
\usepackage{stfloats}
\usepackage{url}
\usepackage{verbatim}
\usepackage{tabularx}  
\usepackage{booktabs}
\usepackage{graphicx}
\usepackage{cite}
\usepackage{csquotes}
\usepackage{xcolor} 
\usepackage{hyperref}
\usepackage[all]{hypcap} 
\renewcommand{\eqref}[1]{\textup{\hyperref[#1]{(\ref*{#1})}}}
\hypersetup{
  colorlinks=true,
  linkcolor=blue,   
  citecolor=blue,   
}
\begin{document}
\bstctlcite{BSTcontrol}
\title{\fontsize{24pt}{28pt}\selectfont Expanding the Transient Stability Region of Attraction of Networked Grid-Interactive Inverters: A Probabilistic Active Learning Framework}

\author{Zhong~Liu,~\IEEEmembership{Student Member,~IEEE},~%
  Jialin~Zheng,~\IEEEmembership{Member,~IEEE},\\
  Junjie~Qin,~\IEEEmembership{Member,~IEEE},~%
  Xiaonan~Lu,~\IEEEmembership{Member,~IEEE}
  
}

\markboth{IEEE trans Smart Grid}%
{Shell \MakeLowercase{\textit{et al.}}: A Sample Article Using IEEEtran.cls for IEEE Journals}


\maketitle

\begin{abstract}
The continuous integration of inverter-based resources makes transient stability analysis increasingly important for power system modernization, in light of the intricate dynamics arising from large-scale inverter deployment. However, analytical transient stability analysis methods consistently yield overly conservative stability boundary estimates, which constrain grid dispatch and operational flexibility. Although recent neural Lyapunov methods attempt to enlarge these stability boundaries to obtain less conservative estimates, they remain trapped within restricted domains due to the out-of-distribution problem. To break this bottleneck, this paper proposes a probabilistic active learning framework. Starting from a deterministic inner stability region certified by neural Lyapunov functions, the framework constructs a Gaussian process surrogate and deploys an uncertainty-guided frontier search. By intelligently coupling electromagnetic transient simulations with active boundary exploration, the algorithm systematically drives the estimated stability boundary outward. Comprehensive evaluations across multi-machine grid-forming benchmarks demonstrate that the proposed framework substantially reduces estimation conservatism. Across test systems ranging up to four interconnected grid-forming inverters, the methodology achieves up to a $20$-fold volumetric enlargement of the certified stability region over classical baselines, while requiring at most 220 time-domain simulation queries per system, far fewer than the 1,600 queries that exhaustive EMT evaluation demands even for the simplest single-inverter benchmark.
\end{abstract}


\section*{Nomenclature}
\addcontentsline{toc}{section}{Nomenclature}
\noindent\textit{Power System Model Symbols}
\begin{IEEEdescription}[\IEEEsetlabelwidth{$T_{\max},\rho_{\max},\varepsilon_{s}$}]
\item[$\delta_{i},\omega_{i}$] Virtual rotor angle and frequency deviation of unit $i$.
\item[$M_{i},H_{i}$] Virtual inertia constant and inertia time constant, $M_{i}=2H_{i}$.
\item[$D_{i}$] Damping coefficient of unit $i$.
\item[$P_{i}^{\ast}$] Active-power setpoint of unit $i$.
\item[$P_{i}^{\mathrm{e}}$] Electrical active power injected by unit $i$.
\item[$E_{i}$] Terminal-voltage magnitude of unit $i$.
\item[$\theta_{\mathrm{COI}},\omega_{\mathrm{COI}}$] COI reference angle and frequency.
\item[$M_{T}$] Total system inertia.
\item[$\Delta\delta_{i},\nu_{i}$] COI-referenced angle and frequency deviations.
\end{IEEEdescription}

\noindent\textit{Gaussian Process and Active Learning Symbols}
\begin{IEEEdescription}[\IEEEsetlabelwidth{$T_{\max},\rho_{\max},\varepsilon_{s}$}]
\item[$\sigma_{f}^{2},\ell$] Signal variance and characteristic length scale.
\item[$\mu(\mathbf{x}),\sigma^{2}(\mathbf{x})$] GP predictive mean and variance.
\item[$\mathbf{K},\mathbf{k}_{\ast}$] Covariance matrix and cross-covariance vector.
\item[$\alpha$] GP regularization parameter.
\item[$\beta,\gamma$] Confidence multiplier and membership threshold.
\item[$J(\mathbf{x})$] UGFS information metric.
\item[$N_{a},N_{m}$] Archive size and number of stochastic probes.
\item[$p_{l},w_{l},q$] Rank-weighted selection probability, weight, and shape parameter.
\item[$\mathbf{b}^{(g)},\boldsymbol{\zeta},\xi$] Adaptive bandwidth, Gaussian noise, and contraction coefficient.
\item[$\varepsilon_{\sigma},d_{\mathrm{div}}$] Uncertainty threshold and minimum query separation.
\end{IEEEdescription}

\section{Introduction}
\IEEEPARstart{T}{he} continuous proliferation of inverter-based resources (IBRs) is driving a profound paradigm shift in the dynamic behavior of modern power grids\cite{intro}. Traditional power systems are dominated by synchronous generators (SGs), whose massive physical rotating inertia acts as a buffer to inherently absorb and suppress energy fluctuations during large-signal disturbances\cite{lowinertia}. As the grid transitions toward a power electronics-dominated paradigm, devices such as Grid-Forming (GFM) inverters are increasingly displacing SGs\cite{xfsg}. Due to their low-inertia or zero-inertia characteristics, these power electronic interfaces respond extremely fast to external grid conditions\cite{11236997, frede}. This implies that when the system encounters severe grid disturbances, the fast control loops of inverters cannot rely on inertia to ride through the transients. Instead, their rapid nonlinear actions may further amplify the initial disturbances, triggering severe transient instability. Consequently, developing reliable transient stability analysis (TSA) methods for such highly nonlinear, low-inertia networks has become an urgent yet unprecedented challenge.

To systematically address the stability challenge, TSA aims to delineate the boundary within which the system can safely recover from severe faults. Mathematically, this transient stability margin is typically quantified by the Region of Attraction (ROA)\cite{zhikang}. In engineering practice, obtaining a non-conservative ROA estimation is of tremendous value\cite{conservative}. In contrast, a highly conservative ROA substantially underestimates the true stability boundary. Given that grid operators generally rely on the estimated boundary as the operational limit for dispatch, an overly conservative ROA artificially shrinks the admissible operating region\cite{hiskens}. Consequently, operators are often forced to adopt suboptimal operational strategies, such as curtailing renewable energy output, executing unnecessary load shedding, or incurring high costs to over-design system hardware capacity\cite{NREL}. Conversely, estimating a less conservative ROA can not only unlock greater transmission potential of existing equipment and higher renewable energy hosting capacity, but also provide broader dispatch headroom for secure grid operation. 

However, traditional TSA methods often suffer from intrinsic theoretical bottlenecks when attempting to expand the ROA  boundary, resulting in extremely conservative evaluation results. Although high-fidelity Electromagnetic Transient (EMT) simulations can provide reliable time-domain testing\cite{EMT}, they can only yield binary stability judgments for limited initial points and cannot explicitly delineate the global ROA boundary in the complete state space. To overcome this limitation, conventional analytical methods attempt to assess transient stability by constructing Lyapunov functions\cite{Lya}. For instance, the Linear Matrix Inequality (LMI) method has to rely on broad Jacobian envelopes to transform the complex inverter dynamics into a convex optimization problem\cite{yuhua}; this excessive approximation directly leads to a severely shrunken estimated ROA. The Sum-of-Squares (SOS) method requires the system dynamics to be in polynomial form, forcing truncation approximations to fit the sinusoidal nonlinear inverter characteristics, which further compresses the provable stability region\cite{SOS}. Due to unavoidable model reduction and structural limitations, these TSA methods consistently yield overly conservative ROA estimates.


Driven by the imperative to accurately uncover these true stability boundaries, recent studies have leveraged neural networks (NNs) to construct neural Lyapunov functions\cite{Tong,my,SCU,Lipschitz}. While these methods mitigate conservativeness to some extent, they ultimately struggle to push the ROA any further toward its physical limits. Early neural Lyapunov frameworks are typically restricted to training within conservative state space regions\cite{Tong,my}. Furthermore, as noted in \cite{chuchu}, when the system dimension exceeds a certain scale, these methods often fail to converge to a valid Lyapunov function.  Subsequent efforts introduced tailored neural Lyapunov candidates to facilitate training\cite{SCU}, but at the cost of limiting the NN's approximation capability, thus introducing additional conservativeness. While recent advancements have successfully achieved rapid algebraic verification using Lipschitz continuity, their estimated ROAs still leave substantial room for improvement regarding the reduction of conservativeness \cite{Lipschitz}. Looking across current NN-based TSA methods, their inability to continuously expand the ROA stems from a common fatal flaw: the Out-of-Distribution (OOD) problem\cite{OOD}. Within the limited training region (i.e., the inner core region around the system equilibrium point), the neural network can satisfy the Lyapunov condition $\dot{V}(x)<0$, where $V(x)$ denotes the NN-parameterized Lyapunov function and $\dot{V}(x)$ is its time derivative. However, once the algorithm attempts to expand the ROA outward into the unknown OOD region, the NNs could suffer from severe generalization errors, causing the gradient $\dot{V}(x)$ to violently distort. This distortion triggers massive "pseudo-counterexamples": state points that are mathematically misjudged as unstable ($\dot{V}(x)\ge0$), but can actually converge to the equilibrium point. These "pseudo-counterexamples" block the possibility of further outward expansion of the ROA, leaving existing NN methods still trapped in the dilemma of conservativeness.

In response to the analytical barriers formed by these "pseudo-counterexamples", this paper proposes a paradigm shift towards a probabilistic active learning framework for TSA. We completely abandon evaluating Lyapunov gradients outside the training region. Instead, inspired by recent advancements in probabilistic modeling \cite{nature}, this work introduces a surrogate modeling philosophy. By leveraging a Gaussian Process (GP) to quantify stability uncertainties and actively querying a high-fidelity EMT Oracle, this data-driven exploration empowers the proposed framework to uncover the less conservative ROAs that remain hidden from traditional analytical tools.

The main contributions of this paper are summarized as follows:
\begin{enumerate}
    \renewcommand{\labelenumi}{\arabic{enumi})}

    \item A three-stage TSA framework is proposed, establishing a structural synergy between rigorous mathematical verification and probabilistic active learning. Specifically, a baseline ROA is first established by a Lipschitz-enforced neural Lyapunov function. Beyond this deterministic inner core, a GP coupled with an active exploration algorithm is deployed to bypass the trap of "pseudo-counterexamples" in the OOD region, thus dynamically extending the stability boundary into the unknown outer space. 
    
    \item The long-standing conservativeness bottlenecks inherent in existing TSA methodologies are substantially mitigated. By bypassing the spurious mathematical violations that hinder NN extrapolation, the proposed methodology captures significantly larger ROAs compared to state-of-the-art baselines. Validated across complex networked GFM inverter test cases, the framework expands the certified ROA size by 5 to 20 times compared to traditional analytical methods. 

    \item High computational and data efficiency is achieved, effectively bridging the gap between pure NNs and pure EMT simulations. Unlike pure NN-based methods that suffer from severe OOD extrapolation errors, or pure EMT approaches that are computationally prohibitive for delineating continuous stability boundaries, the proposed framework synthesizes a reliable, less conservative ROA while demanding only a marginal fraction of EMT runs. 

\end{enumerate}

The remainder of this paper is structured as follows. Section II establishes the GFM inverter system modeling. Section III details the GP-based active exploration methodology. Section IV provides the case studies. Section V draws the conclusions.

\section{Preliminary: Modeling of Networked Grid-Interactive Inverters}
To endow the inverter-dominated grid with the inertial and damping response, this work adopts GFM units as a representative class of IBRs. Each GFM unit is operated under virtual synchronous machine (VSM) control, which reproduces the electromechanical behavior of a SG within the inverter control loop \cite{ncepu}. For the $i$-th unit in a network of $N$ interconnected units, the active power loop yields the swing dynamics
\begin{align}
\dot{\delta}_{i} &= \omega_{i}, \label{eq:swing1}\\
M_{i}\,\dot{\omega}_{i} &= P_{i}^{\ast} - P_{i}^{\mathrm{e}} - D_{i}\,\omega_{i}, \label{eq:swing2}
\end{align}
where $\delta_{i}$ and $\omega_{i}$ are the virtual rotor angle and the angular-frequency deviation of unit $i$, $M_{i}$ is the virtual inertia constant, $D_{i}$ is the damping coefficient, and $P_{i}^{\ast}$ is the active-power setpoint.

These GFM inverters are electrically coupled through the impedance network; consequently, the active power injected by unit $i$ is governed by the relative virtual angles of its interconnected neighbors, formulated as:
\begin{equation}
P_{i}^{\mathrm{e}} = \sum_{j=1}^{N} E_{i}E_{j}\bigl( G_{ij}\cos\delta_{ij} + B_{ij}\sin\delta_{ij} \bigr),\quad \delta_{ij}=\delta_{i}-\delta_{j},
\label{eq:Pe}
\end{equation}
where $E_{i}$ is the terminal-voltage magnitude of unit $i$, and $G_{ij}$ and $B_{ij}$ are the real and imaginary parts of the $(i,j)$ entry of the network admittance matrix. Because the reactive-power loop holds each voltage magnitude close to its reference, $E_{i}$ is treated as constant, so the transient stability of the networked system is governed by the coupled angle--frequency dynamics in \eqref{eq:swing1}--\eqref{eq:Pe}.

The post-fault steady state ${\delta}^{\ast}$ is the equilibrium point of \eqref{eq:swing1}--\eqref{eq:Pe}, obtained by setting the time derivatives to zero. However, a significant challenge arises when directly analyzing the system in the absolute reference frame. Since the electrical power injection $P_{i}^{\mathrm{e}}$ is exclusively dictated by the relative angle differences $\delta_{ij}$, shifting all absolute virtual angles by an identical arbitrary constant yields zero effect on the system dynamics. This physical characteristic implies that the post-fault equilibrium is not a single isolated point, but rather a continuous infinite manifold. Since rigorous Lyapunov-based asymptotic stability certification strictly demands an isolated equilibrium point\cite{nonlinear}, this absolute reference frame completely hinders direct Lyapunov-based TSA.

To overcome this issue, the system dynamics are mapped into the center-of-inertia (COI) reference frame \cite{sauer}. In this reference frame, each unit's angle and frequency are measured relative to the inertia-weighted dynamic average of the entire network. Consequently, the state variables of one unit can be algebraically eliminated using the COI constraints, as detailed in Appendix~\ref{app:coi}. By shifting the newly anchored post-fault equilibrium point to the origin, the reduced dynamics successfully form an autonomous system. The state vector of this reduced system is defined as:
\begin{equation}
    \mathbf{x} = \bigl[\Delta\delta_{1},\dots,\Delta\delta_{N-1},\,\nu_{1},\dots,\nu_{N-1}\bigr]^{\top} \in \mathbb{R}^{n}, \label{eq:state_vector}
\end{equation}
where $n=2(N-1)$. This vector stacks the COI-referenced angle deviations $\Delta\delta_{i}$ and frequency deviations $\nu_{i}$ of the $N-1$ independent units. The post-fault equilibrium point is now an isolated, asymptotically stable equilibrium, which serves as the prerequisite for the proposed method in Section III.

Building upon the formulated autonomous system, the transient stability of the networked GFM inverters can be analytically evaluated via Lyapunov's direct method. The essence of this classical theorem lies in constructing a continuously differentiable, positive-definite scalar function $V(\mathbf{x})$. The system equilibrium point is guaranteed to be asymptotically stable if the time derivative of $V(\mathbf{x})$ is strictly negative:
\begin{equation}
    \mathcal{L}_{f} V(\mathbf{x}) = \nabla V(\mathbf{x}) \cdot f(\mathbf{x}) < 0.
    \label{lie}
\end{equation}
\begin{figure}
    \centering
    \includegraphics[width=1\linewidth]{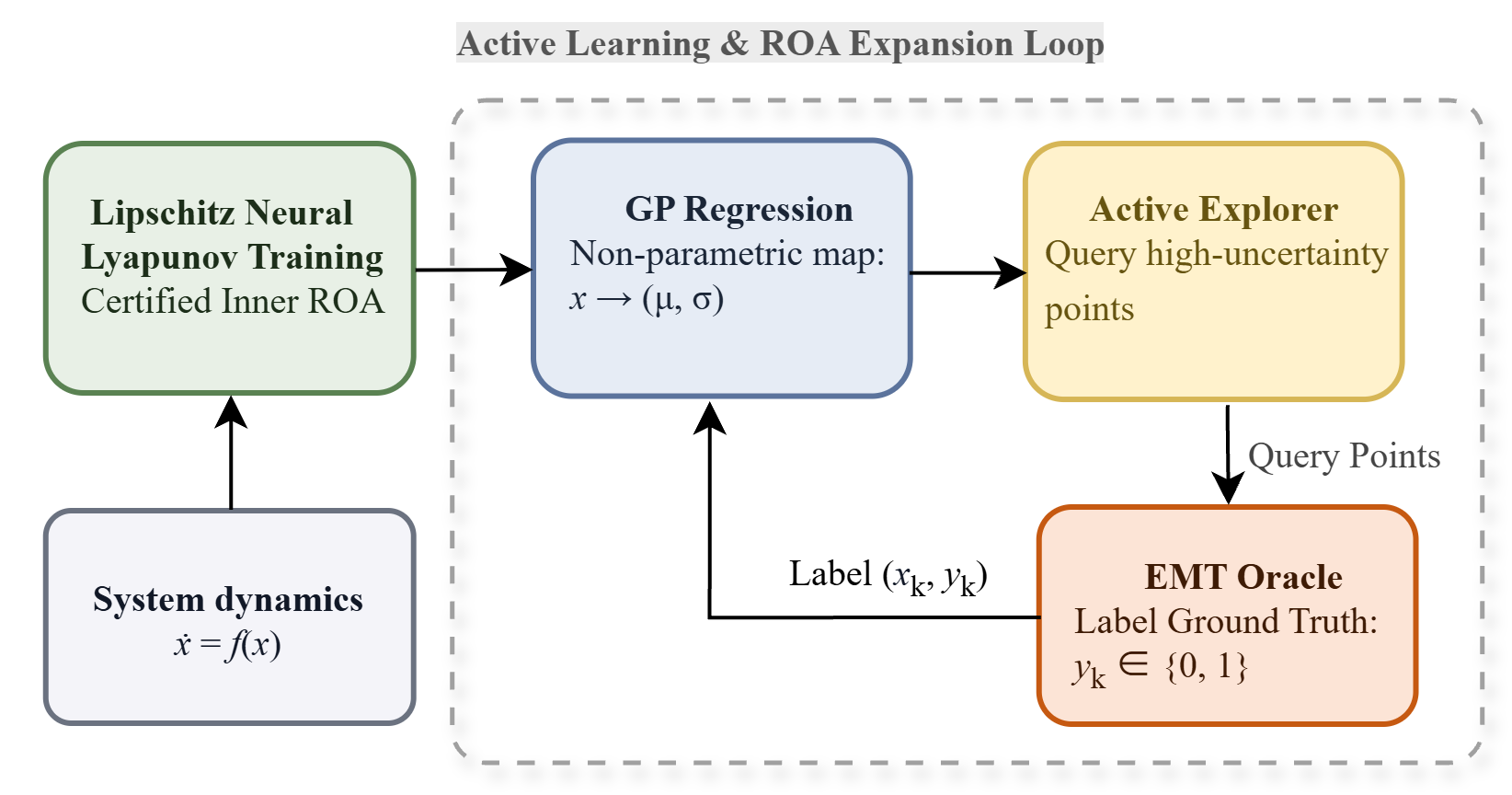}
    \caption{Overall architecture of the proposed framework.}
    \label{fig:schematic}
    \vspace{-0.8cm}
\end{figure}

\vspace{-1cm}
\section{Methodology: Probabilistic Active Learning Framework}
\subsection{Framework Overview}
As illustrated in Fig.~\ref{fig:schematic}, the proposed methodology adopts a synergistic architecture comprising a deterministic mathematical anchor and a probabilistic active learning loop. Rooted in the analytical system dynamics, the framework first executes a Lipschitz-enforced neural Lyapunov training. This foundational process establishes a rigorously verified inner ROA, yielding a deterministic baseline for transient stability (detailed in Section III-B).

To push the stability boundary beyond the conservative limits of this Lyapunov certificate, the framework transitions into the Active Learning and ROA Expansion Loop. Within this closed-loop architecture, a GP regression is first constructed as a non-parametric surrogate of the system's stability indicator (Section III-C). By mapping the state space to a probabilistic stability assessment, the GP outputs both the predictive mean and its associated uncertainty. Driven by this uncertainty quantification, an active explorer known as Uncertainty-Guided Frontier Search (UGFS) is deployed to target informative areas along the stability frontier (Section III-D). Specifically, the explorer samples state-space locations where the current predictive uncertainty is highest. By invoking the EMT Oracle at these targeted points, the framework maximizes the subsequent reduction of uncertainty once the new data is acquired. The Oracle evaluates the actual transient responses to generate discrete ground-truth labels, which are subsequently fed back to refine the GP regression. This iterative closed-loop interaction rapidly collapses the predictive uncertainty, effectively driving the continuous expansion of the ROA into previously unexplored regions.

Throughout this section, the post-fault dynamics of the networked GFM inverter system established in Section II are written compactly as
\begin{equation}
\dot{\mathbf{x}} = f(\mathbf{x}), \qquad \mathbf{x}\in\mathbb{R}^{n},
\label{eq:sys}
\end{equation}
where the state vector has been shifted such that the post-fault stable equilibrium point (SEP) coincides with the origin, i.e., $f(\mathbf{0})=\mathbf{0}$. The true (yet unknown) ROA is denoted by
\begin{equation}
\mathcal{R}^{\star} := \bigl\{\mathbf{x}_{0}\in\mathbb{R}^{n} : \lim_{t\to\infty}\mathbf{x}(t;\mathbf{x}_{0}) = \mathbf{0}\bigr\},
\label{eq:trueroa}
\end{equation}
where $\mathbf{x}(t;\mathbf{x}_{0})$ is the solution of \eqref{eq:sys} initialized at $\mathbf{x}_{0}$.
\begin{figure}[t]
    \centering
    \includegraphics[width=0.75\linewidth]{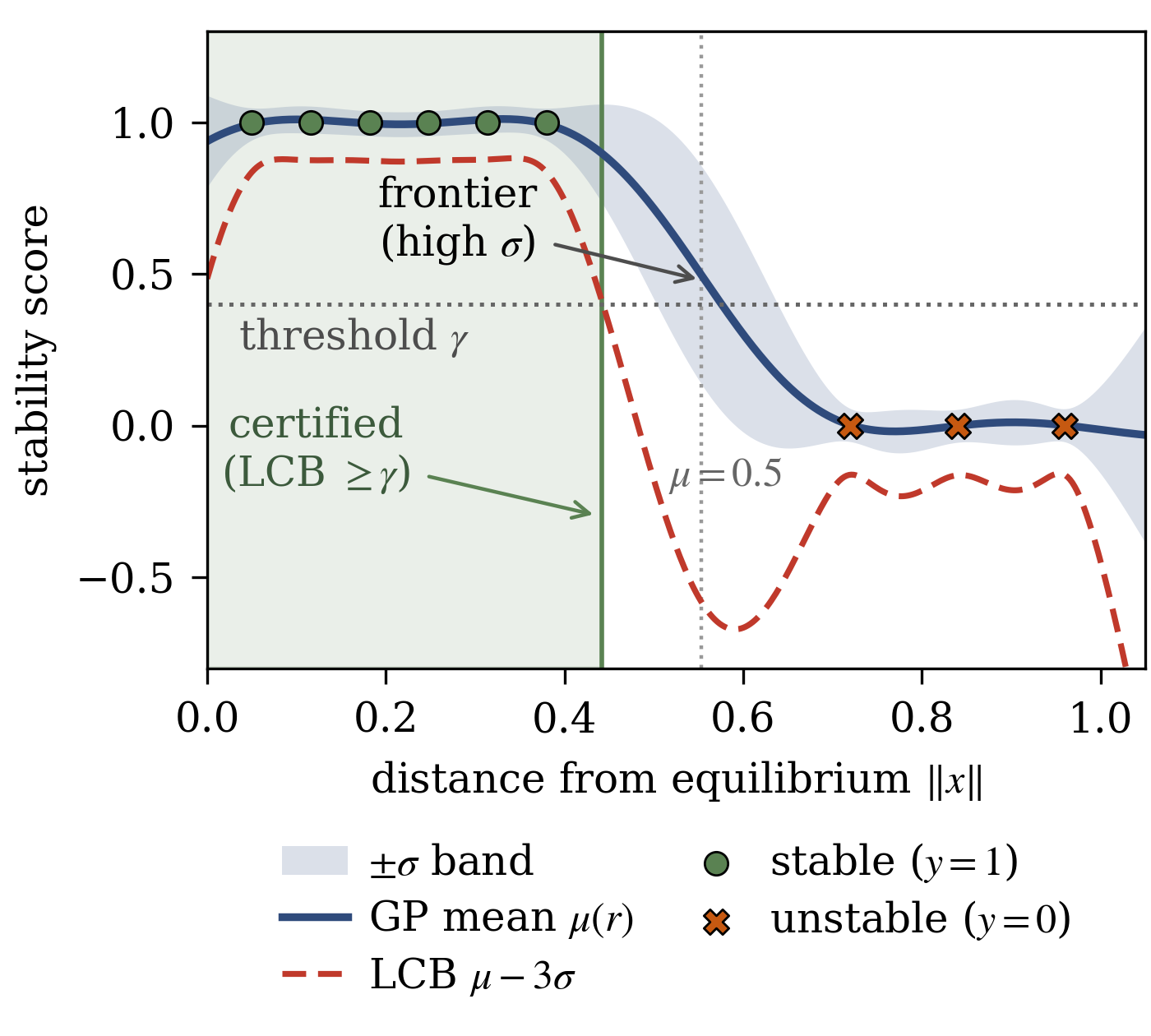}
    \vspace{-0.5cm}
    \caption{One-dimensional cross-section illustrating GP regression of the stability boundary.}
    \label{fig:gp}
    \vspace{-0.7cm}
\end{figure}
\subsection{Lipschitz-Certified Construction of the Base ROA} \label{subsec:stage A}

The first stage anchors the entire active learning framework with a foundational ROA rigorously validated by mathematical deduction. Following the methodology in \cite{Lipschitz}, a neural Lyapunov candidate $V_{\theta}(\mathbf{x})$ is parameterized by a feedforward neural network. The network architecture is explicitly structured with a squared-norm output layer, ensuring the positive-definite condition $V_{\theta}(\mathbf{x})\ge 0$ by construction. Consequently, the training phase focuses solely on enforcing the negativity of the Lie derivative $\mathcal{L}_{f}V_{\theta}(\mathbf{x})<0$ over a verification ball $\mathcal{B}_{R}:=\{\mathbf{x}:\|\mathbf{x}\|_{2}\le R\}$.

However, empirical training over discrete collocation points inherently lacks continuous-domain guarantees. Instead of relying on computationally intensive Satisfiability Modulo Theories (SMT) solvers, an algebraic verification approach rooted in Lipschitz continuity is adopted to certify the empirical candidate. Specifically, the continuous domain $\mathcal{B}_{R}$ is discretized into a dense verification lattice $\mathcal{X}_{\tau}$ with a covering radius $\tau$. Let $L_{f}$ and $L_{V}$ denote the Lipschitz constants of the system vector field $f$ and the gradient of the Lyapunov candidate $\nabla V_{\theta}$, respectively. By enforcing a stricter margin on the discrete lattice points, the pointwise condition:
\begin{equation}
\mathcal{L}_{f}V_{\theta}(\mathbf{x}_{k}) < -L_{V}\bigl(1+L_{f}\bigr)\,\tau, \qquad \forall\, \mathbf{x}_{k}\in\mathcal{X}_{\tau},
\label{eq:cert}
\end{equation}
algebraically guarantees that $\mathcal{L}_{f}V_{\theta}(\mathbf{x})<0$ holds continuously for all $\mathbf{x}\in\mathcal{B}_{R}$. 

Once the NN is successfully verified across the entire lattice, the certified base ROA is extracted as the largest positively invariant sublevel set fully contained within the verified ball:
\begin{equation}
c^{\ast} = \min_{\|\mathbf{x}\|_{2}=R} V_{\theta}(\mathbf{x}), \qquad
\Omega_{\mathrm{base}} = \bigl\{\mathbf{x}\in\mathcal{B}_{R} : V_{\theta}(\mathbf{x})\le c^{\ast}\bigr\}.
\label{eq:base}
\end{equation}
By classical Lyapunov arguments, any post-fault state strictly originating within $\Omega_{\mathrm{base}}$ is deterministically guaranteed to converge to the stable equilibrium. 


\subsection{GP-Based Probabilistic Expansion of the Stability Boundary}\label{subsec:stage B}

Beyond the mathematically certified limits of $\Omega_{\mathrm{base}}$, the proposed framework abandons analytical Lyapunov gradients to circumvent the aforementioned OOD distortion. Instead, the TSA is reformulated to rely directly on the definitive physical outcomes provided by EMT simulations. Specifically, for any post-fault state $\mathbf{x}$, a binary physical indicator representing whether the system successfully synchronizes or completely diverges is provided:
\begin{equation}
y(\mathbf{x}) = \mathbf{1}\{\mathbf{x}\in\mathcal{R}^{\star}\} \in \{0,1\},
\label{eq:indicator}
\end{equation}

Since acquiring this ground-truth label $y(\mathbf{x})$ demands a computationally expensive time-domain simulation, exhaustive evaluation across the continuous state space is prohibitive. Consequently, a robust surrogate model is deployed to interpolate stability predictions from sparse EMT observations while rigorously quantifying the associated uncertainty. To this end, a GP prior is placed on a latent stability field $g(\mathbf{x})\sim\mathcal{GP}\bigl(0, k(\mathbf{x},\mathbf{x}')\bigr)$, utilizing a standard squared-exponential kernel to capture the smooth physical transitions within the state space:
\begin{equation}
k(\mathbf{x},\mathbf{x}') = \sigma_{f}^{2}\exp\!\Bigl(-\frac{\|\mathbf{x}-\mathbf{x}'\|_{2}^{2}}{2\ell^{2}}\Bigr),
\label{eq:kernel}
\end{equation}
where $\sigma_{f}^{2}$ is the signal variance and $\ell$ denotes the characteristic length scale. Given a training dataset $\mathcal{D}=\{(\mathbf{x}_{i}, y_{i})\}_{i=1}^{M}$ with labels $y_{i}\in\{0,1\}$, the GP yields a Gaussian posterior at any arbitrary test state $\mathbf{x}$. This posterior provides two critical metrics, a predictive mean $\mu(\mathbf{x})$ and a predictive variance $\sigma^{2}(\mathbf{x})$:
\begin{align}
\mu(\mathbf{x}) &= \mathbf{k}_{\ast}^{\top}\bigl(\mathbf{K}+\alpha\mathbf{I}\bigr)^{-1}\mathbf{y}, \label{eq:gpmean}\\
\sigma^{2}(\mathbf{x}) &= k(\mathbf{x},\mathbf{x}) - \mathbf{k}_{\ast}^{\top}\bigl(\mathbf{K}+\alpha\mathbf{I}\bigr)^{-1}\mathbf{k}_{\ast}. \label{eq:gpvar}
\end{align}
Here, $[\mathbf{K}]_{ij}=k(\mathbf{x}_{i},\mathbf{x}_{j})$ and $[\mathbf{k}_{\ast}]_{i}=k(\mathbf{x}_{i},\mathbf{x})$. The regularization parameter $\alpha>0$ plays a pivotal dual role: it ensures the numerical well-conditioning of the covariance matrix and acts as a soft-margin, acknowledging that early prior labels can be gracefully overruled by subsequent EMT evidence. The hyperparameters $(\sigma_{f},\ell)$ are continuously re-optimized via the log marginal likelihood, allowing the GP's spatial resolution to automatically adapt to the underlying geometry of the expanding stability boundary:
\begin{equation}
\log p(\mathbf{y}\,|\,\mathcal{D}) = -\tfrac{1}{2}\mathbf{y}^{\top}\widetilde{\mathbf{K}}^{-1}\mathbf{y} - \tfrac{1}{2}\log\bigl|\widetilde{\mathbf{K}}\bigr| - \tfrac{M}{2}\log 2\pi,
\label{eq:mll}
\end{equation}
with $\widetilde{\mathbf{K}}=\mathbf{K}+\alpha\mathbf{I}$.

To establish a geometric intuition for these concepts, Fig.~\ref{fig:gp} illustrates a one-dimensional cross-section of the state space extending outward from the equilibrium. The solid blue curve represents the predictive mean $\mu(\mathbf{x})$, translating directly to the \emph{Bayesian probability} of transient stability (decaying from $1$ near the origin to $0$ in the unstable far field). The shaded blue region denotes the uncertainty $\pm\sigma(\mathbf{x})$, dynamically bulging in the unexplored gaps where labeled data is absent. Subtracting this uncertainty from the mean yields the Lower Confidence Bound (LCB, dashed red line), representing a high-probability lower bound on stability. Crucially, a state is admitted into the certified probabilistic ROA (the green shaded region) only if its LCB strictly clears the predefined membership threshold $\gamma$. By enforcing this geometric condition, the predictive uncertainty natively acts as a built-in safety margin, ensuring the certified boundary remains confined within the true physical limits until targeted EMT queries locally collapse the uncertainty band.

\subsubsection{GP Surrogate Initialization}A significant computational advantage of the proposed framework is the capability to establish an initial probabilistic surrogate of the system without executing EMT simulations. This initialization leverages established physical priors inherent to power system dynamics. Specifically, the interior base ROA is already certified; any state $\mathbf{x}_{i}$ encapsulated within the baseline ROA, $\Omega_{\mathrm{base}}$, intrinsically guarantees transient stability. Consequently, positive stability labels $\{(\mathbf{x}_{i},1)\}$ can be extracted from the baseline ROA without additional computational overhead. As illustrated in Fig.~\ref{fig:gp}, these initial positive samples are depicted as the stable points ($y=1$, denoted by green circles).

Conversely, extreme state deviations inevitably cause an irreversible loss of synchronism. This physical characteristic is encoded by allocating negative pseudo-labels $\{(\mathbf{x}_{j},0)\}$ within the far-field regions of the state space, represented by the unstable markers ($y=0$, denoted by orange crosses). Facilitated by the soft-label parameter $\alpha$, these negative priors remain adaptable rather than permanently fixed. If active exploration later proves a far-field state is actually stable, the GP regression updates the local predictive mean. The resulting initial dataset, $\mathcal{D}_{0}$, perfectly incorporates the definitively stable labels and the inherently unstable labels, leaving a wide, highly uncertain "frontier" in the middle (annotated in Fig.~\ref{fig:gp}) that the algorithm must actively explore.
\subsubsection{EMT Oracle and Trajectory Inheritance}
To explore the high-uncertainty frontier, the ground-truth EMT Oracle $\mathcal{O}$ is actively queried. For any specific initial state $\mathbf{x}_{0}$, the Oracle executes a time-domain simulation. If the system regains stability within a defined time horizon $T_{\max}$ without exceeding the divergence threshold $\rho_{\max}$, a stable verdict is returned:
\begin{equation}
y = \mathbf{1}\Bigl\{ \sup_{t\le T_{\max}}\|\mathbf{x}(t)\|_{2} < \rho_{\max} \;\wedge\; \|\mathbf{x}(T_{\max})\|_{2} < \varepsilon_{s} \Bigr\}.
\label{eq:oracle}
\end{equation}
Given that EMT simulations are computationally heavy, relying on them to label single initial states is highly inefficient. From a dynamical systems perspective, if the system successfully converges to the equilibrium from an initial state $\mathbf{x}_{0}$, the entire traversed transient trajectory inherently resides within the true ROA \cite{nonlinear}. Consequently, whenever a stable verdict ($y=1$) is obtained, $N_{\mathrm{traj}}$ points are subsampled along this convergent path and incorporated into the GP training dataset as guaranteed positive labels:
\begin{equation}
\mathcal{D} \leftarrow \mathcal{D} \cup \{(\mathbf{x}_{0},y)\} \cup \bigl\{(\mathbf{x}(t_{j}),1)\bigr\}_{j=1}^{N_{\mathrm{traj}}}, \quad \text{if } y=1.
\label{eq:inherit}
\end{equation}

\begin{figure}[t]
    \centering
    \includegraphics[width=0.75\linewidth]{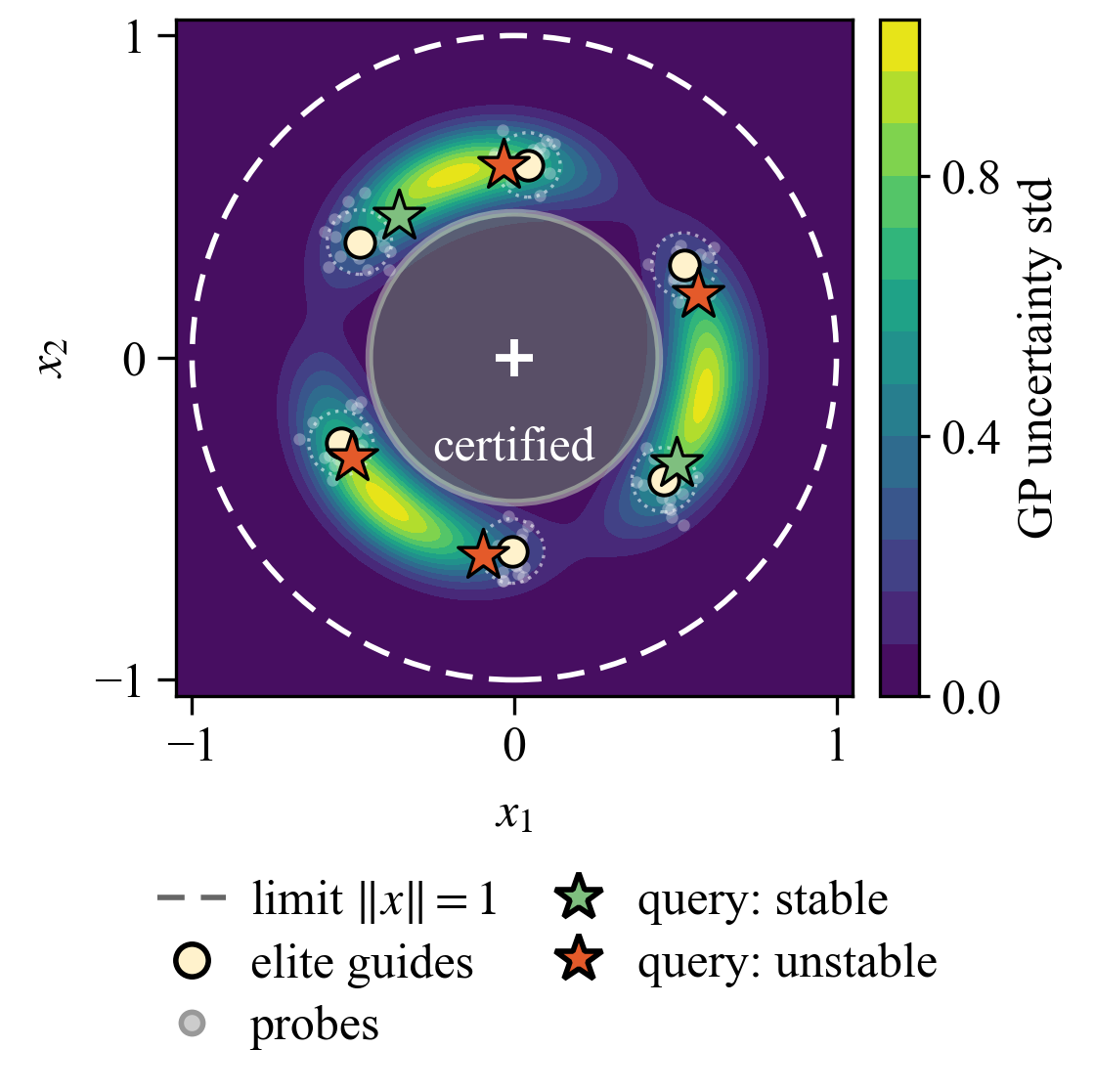}
    \vspace{-0.5cm}
    \caption{Two-dimensional COI state-space projection illustrating the UGFS boundary exploration mechanism.}
    \label{fig:explorer}
    \vspace{-0.5cm}
\end{figure}
\begin{figure*}
    \centering
    \includegraphics[width=1\linewidth]{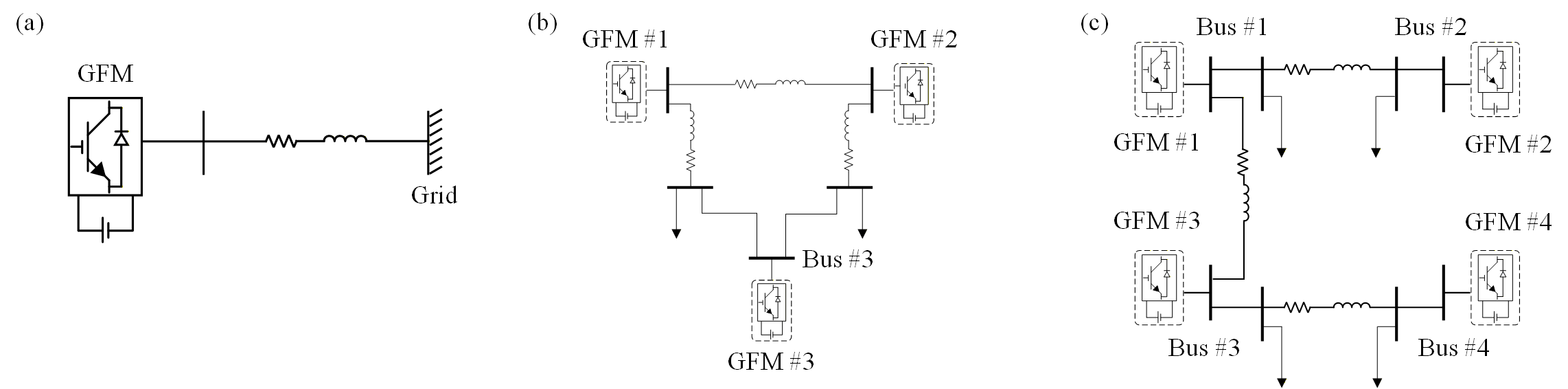}
    \caption{Schematics of the three test systems: (a) single-GFM infinite-bus system. (b) 3-GFM system. (c) 4-GFM system.}
    \label{3cases}
    \vspace{-0.6cm}
\end{figure*}
\subsubsection{Probabilistic ROA Readout}
Applying the LCB criterion of Fig.~\ref{fig:gp} across the search domain $\Omega_{\mathrm{sr}}$ yields the certified probabilistic ROA:
\begin{equation}
\widehat{\mathcal{R}} = \mathcal{C}_{0}\Bigl( \bigl\{\mathbf{x}\in\Omega_{\mathrm{sr}} : \mu(\mathbf{x}) - \beta\,\sigma(\mathbf{x}) \ge \gamma \bigr\} \Bigr) \cup \Omega_{\mathrm{base}},
\label{eq:final}
\end{equation}
where $\beta$ is the confidence multiplier, $\gamma$ is the membership threshold, and $\mathcal{C}_{0}(\cdot)$ retains only the connected component anchored at $\Omega_{\mathrm{base}}$.
\subsection{Uncertainty-Guided Frontier Search for ROA expansion}\label{subsec:stage C}
To systematically estimate a less conservative and enlarged ROA, the Uncertainty-Guided Frontier Search (UGFS) is introduced. Rather than employing uniform or passive sampling, UGFS operates as a dynamic boundary-tracking mechanism designed to concentrate computational resources on actively expanding the established stability boundary. As illustrated in the two-dimensional state-space projection in Fig.~\ref{fig:explorer}, the predictive uncertainty $\sigma(\mathbf{x})$ forms a narrow, high-variance shell surrounding the certified ROA (the central disk). UGFS is specifically formulated to strategically target and sample this unexplored boundary.

To evaluate candidate states, UGFS utilizes an information metric:
\begin{equation}
J(\mathbf{x}) = \sigma(\mathbf{x})\cdot \mathbf{1}\bigl\{\mu(\mathbf{x}) > \tfrac{1}{2}\bigr\},
\label{eq:fitness}
\end{equation}
This formulation is designed to maximize information gain while mitigating the misallocation of computational resources. The predictive uncertainty $\sigma(\mathbf{x})$ naturally peaks in the transitional regions where labeled EMT data are sparse. However, blindly exploring all high-uncertainty areas would waste time-intensive simulations on extreme far-field states where a loss of physical synchronism is already practically guaranteed. To circumvent this, the indicator function $\mathbf{1}\{\mu(\mathbf{x}) > 1/2\}$ operates as a probabilistic heuristic filter. It explicitly nullifies the exploration value of any state that the surrogate model predicts as likely unstable. 

Consequently, the algorithm is constrained to advance the stability boundary from the inside out. It exclusively queries highly uncertain states that still maintain a realistic probability of surviving the transient. As visualized in Fig.~\ref{fig:explorer}, this logical condition restricts the high-value candidates to the inner, more secure perimeter of the high-uncertainty ring.

The maximization of \eqref{eq:fitness} is executed through an iterative stochastic search guided by a discrete set of highly informative boundary states. Specifically, an elite archive $\mathcal{A}=\{\mathbf{x}_{(1)},\dots,\mathbf{x}_{(N_{a})}\}$ is dynamically updated to retain the $N_{a}$ highest-scoring candidate states evaluated throughout the search progression, sorted in descending order of $J(\mathbf{x})$. These archived states function as spatial guide markers distributed along the high-uncertainty stability frontier, as depicted in Fig.~\ref{fig:explorer}. During each internal algorithmic iteration, $N_{m}$ stochastic state probes are generated. To instantiate a new probe, a discrete guide $\mathbf{x}_{(l)}$ is first selected from the archive utilizing a rank-weighted probability distribution:
\begin{equation}
p_{l} = \frac{w_{l}}{\sum_{j=1}^{N_{a}} w_{j}}, \qquad
w_{l} = \frac{1}{qN_{a}\sqrt{2\pi}}\exp\!\Bigl(-\frac{(l-1)^{2}}{2q^{2}N_{a}^{2}}\Bigr),
\label{eq:rankweight}
\end{equation}
where the parameter $q$ explicitly dictates the shape of this Gaussian probability curve over the rankings. A smaller $q$ narrows the curve, heavily concentrating the selection probability on the few highest-ranked guides. In contrast, a larger $q$ flattens the curve, yielding a more uniform selection probability across the entire archive. Once a specific guide is selected, a new candidate state is generated:
\begin{equation}
\mathbf{x}' = \mathbf{x}_{(g)} + \mathbf{b}^{(g)}\odot\boldsymbol{\zeta}, \qquad \boldsymbol{\zeta}\sim\mathcal{N}(\mathbf{0},\mathbf{I}_{n}),
\label{eq:probe}
\end{equation}
To automatically adjust the search resolution, the spatial variance of this Gaussian kernel is dynamically scaled based on the local clustering of the archive points around the selected guide:
\begin{equation}
b^{(g)}_{d} = \xi \sum_{l=1}^{N_{a}} \frac{\bigl| x_{(l),d} - x_{(g),d} \bigr|}{N_{a}-1}, \qquad d = 1,\dots,n,
\label{eq:bandwidth}
\end{equation}
where $\xi\in(0,1)$ is a contraction coefficient. Probes instantiated outside the predefined physical search limits $\Omega_{\mathrm{sr}}$ are radially projected back into the valid operational domain. After evaluation via \eqref{eq:fitness}, the combined set of archived states and new probes is truncated to retain only the $N_{a}$ optimal members. Any probe exhibiting a mathematically significant uncertainty, $J(\mathbf{x}')\ge\varepsilon_{\sigma}$, is recorded into an aggregate candidate set $\mathcal{H}$, which serves as the designated candidate pool for subsequent EMT simulations. Visually, each guide marker in Fig.~\ref{fig:explorer} is thus enveloped by a localized cloud of stochastic probes, with the adaptive kernel width represented by the dotted ellipse.


Following the search phase, the final EMT simulation queries are extracted from the merged candidate set $\mathcal{H}$. To select the $N_{b}$ allocated queries, candidates are evaluated in descending order of their information metric $J(\mathbf{x})$ and admitted only if they maintain a minimum Euclidean distance $d_{\mathrm{div}}$ from all previously selected points. This minimum distance constraint ensures that time-intensive EMT simulations are well-distributed along the expanding stability boundary rather than clustering in one localized area. The selected states are then evaluated by the time-domain oracle \eqref{eq:oracle}, the training dataset is augmented via trajectory inheritance \eqref{eq:inherit}, and the GP posterior is refitted. Consequently, the predictive uncertainty landscape updates, and the certified boundary migrates outward wherever stable verdicts are confirmed. As illustrated in Fig.~\ref{fig:explorer}, a stable verdict (green star) pushes the certified ROA outward along that specific direction, whereas an unstable verdict (orange star) identifies the true power system stability limit.

This cyclic alternation between boundary identification and GP refitting proceeds until the uncertainty collapse criterion is satisfied:
\begin{equation}
\max_{\mathbf{x}\in\mathcal{H}} J(\mathbf{x}) < \varepsilon_{\sigma} \quad \text{for } Q_{s} \text{ consecutive rounds}.
\label{eq:stop}
\end{equation}
This condition indicates that wherever the system is predicted to survive ($\mu > 1/2$), the surrogate model is now highly confident ($\sigma < \varepsilon_{\sigma}$). At this juncture, the surrogate model has robustly approximated the physical stability limits, and the final extended ROA $\widehat{\mathcal{R}}$ is extracted.

\section{Case Studies}
The proposed framework is evaluated on three benchmark systems of increasing dimensionality: a single-machine infinite-bus (SMIB) system ($n=2$), the three-GFM system in the COI frame ($n=4$), and a four-GFM system formulated in the COI frame ($n=6$). In all cases, the EMT oracle \eqref{eq:oracle} is implemented as a full nonlinear time-domain integration of the post-fault dynamics with an adaptive Runge--Kutta solver, a divergence threshold $\rho_{\max}=5$~p.u., and a settling tolerance $\varepsilon_{s}=0.05$. The probabilistic readout uses $\beta=3$ and $\gamma=0.9$ throughout, and the UGFS is configured identically across all cases: $N_{a}=50$, $N_{m}=40$, $q=0.1$, $\xi=0.85$, $\varepsilon_{\sigma}=0.05$, and $d_{\mathrm{div}}=0.15$. The estimates are benchmarked against the Lipschitz-certified ROA and a quadratic Lyapunov certificate obtained from the LQR method.

\begin{figure}
\centering
\includegraphics[width=1\linewidth]{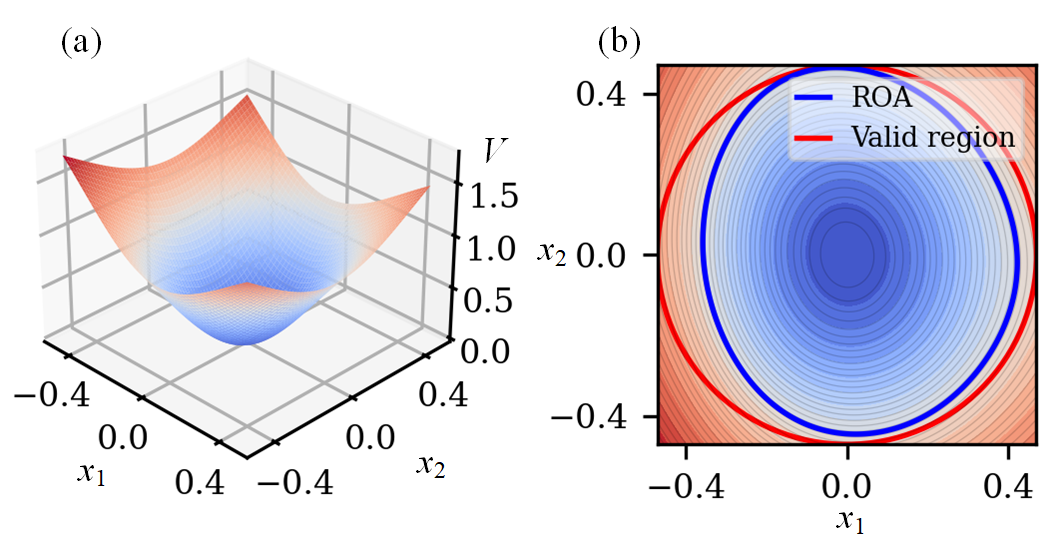}
\caption{Training results: (a) neural Lyapunov functions of single-GFM infinite bus system. (b) ROA estimated by Lipschitz-enforced method in \cite{Lipschitz}.}
\label{fig:roa_concept}
\end{figure}
\begin{figure}
\centering
\includegraphics[width=0.8\linewidth]{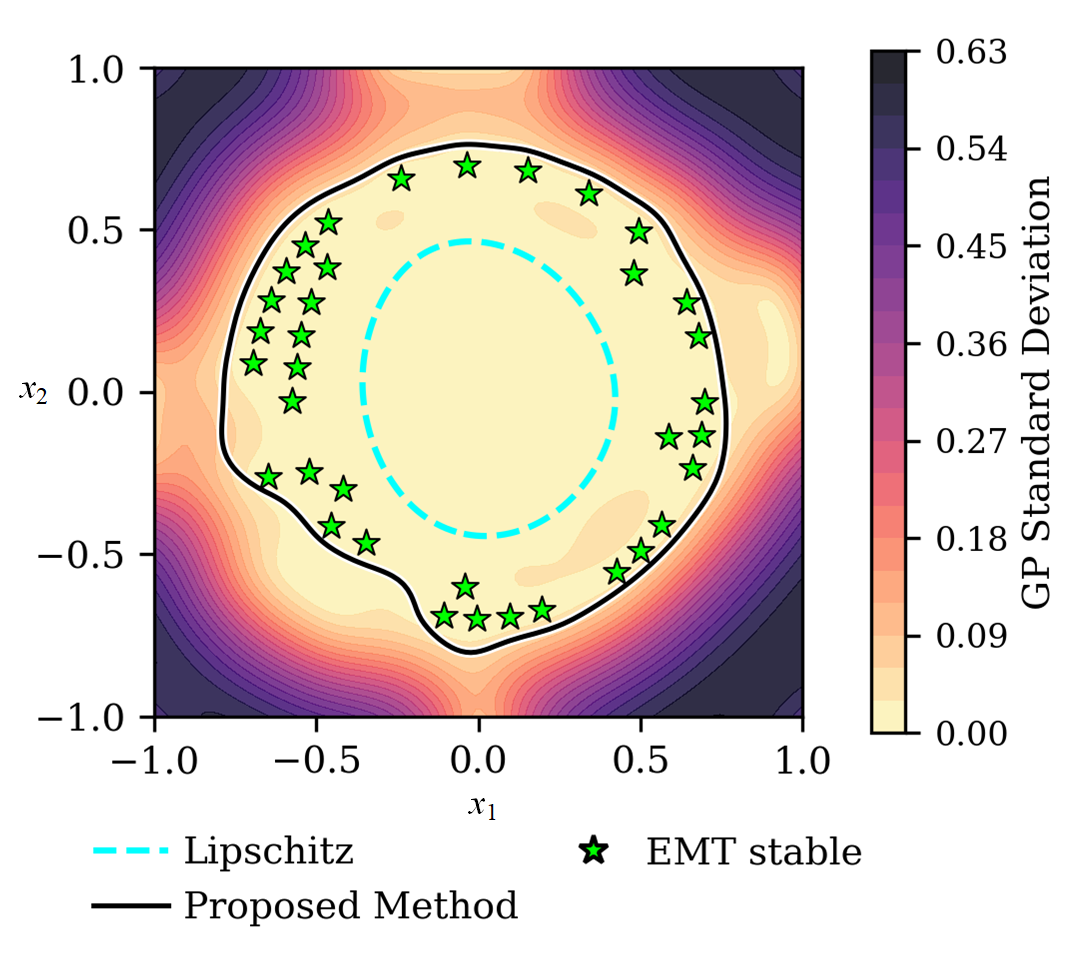}
\vspace{-0.5cm}
\caption{Probabilistic ROA expansion for Case I in the $(x_1,x_2)$ state space.}
\label{fig:framework}
\end{figure}

\subsection{Case I: Single GFM System}
The first benchmark is the single GFM configuration (Fig. \ref{3cases}\,(a)) whose post-fault dynamics in SEP-shifted coordinates $\mathbf{x}=[x_{1},x_{2}]^{\top}$ read
\begin{equation}
\dot{x}_{1} = x_{2}, \qquad
\dot{x}_{2} = \frac{1}{M}\bigl(P_{m} - P_{\max}\sin(x_{1}+\delta_{s}) - Dx_{2}\bigr),
\label{eq:smib}
\end{equation}
with $M=1.0$, $D=0.1$, $P_{m}=2.0$, $P_{\max}=2.5$ (all in p.u.), and $\delta_{s}=\arcsin(P_{m}/P_{\max})$.

In the first stage (\ref{subsec:stage A}), the baseline analytical certificate is constructed by training a Lipschitz-enforced neural Lyapunov network configured with 16 neurons per hidden layer over a training domain $\mathcal{B}_{R}$ with $R=0.5$. As illustrated in Fig.~\ref{fig:roa_concept}, the trained neural network successfully establishes a rigorous mathematical foundation. Fig.~\ref{fig:roa_concept}\,(a) shows the learned Lyapunov function $V(\mathbf{x})$, forming a smooth, positive-definite energy landscape around the equilibrium. Fig.~\ref{fig:roa_concept}\,(b) delineates the ROA of $V(\mathbf{x})$. While mathematically rigorous, this analytical boundary exhibits notable conservatism relative to the physical stability limit of the inverter.

Following the initialization in \ref{subsec:stage B} and the execution of the UGFS in \ref{subsec:stage C}, the final probabilistic stability map is generated. Fig.~\ref{fig:framework} displays the expanded stability boundary, whose background contour visualizes the GP predictive standard deviation $\sigma(\mathbf{x})$. The proposed method substantially enlarges the conservative Lipschitz-certified core. This outward expansion is systematically driven by informative EMT simulation queries (green stars) deposited along the advancing high-uncertainty ridge.

\begin{figure}
\centering
\includegraphics[width=1\linewidth]{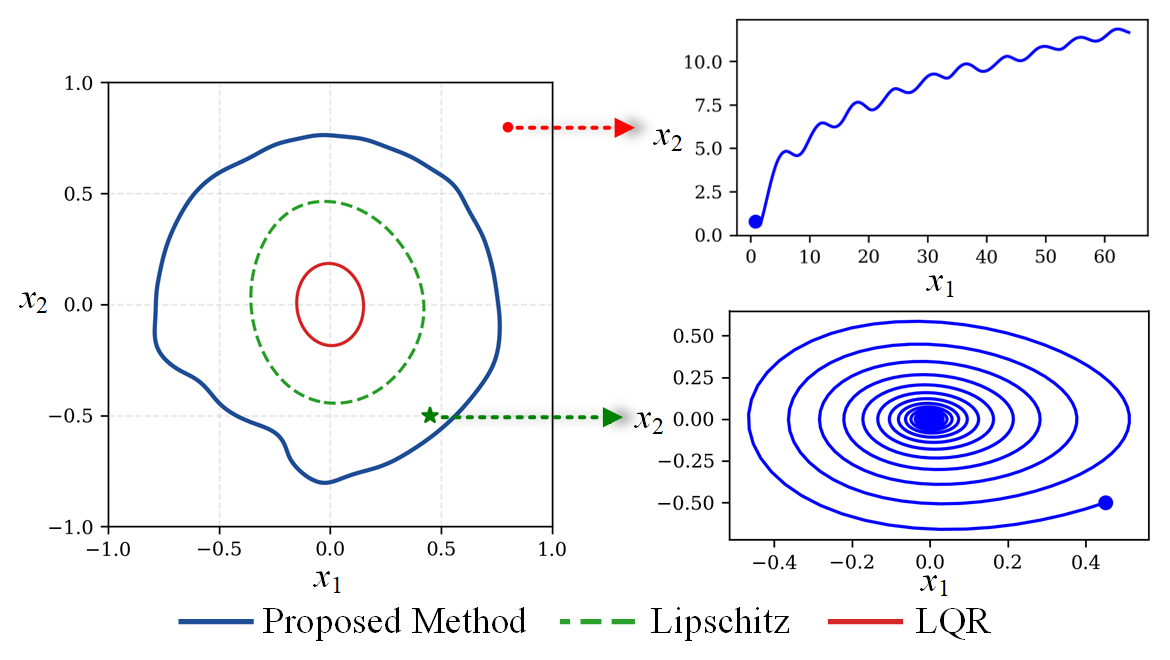}
\caption{ROA comparison and time-domain verification for Case I.}
\label{fig:roa_9bus}
\vspace{-0.3cm}
\end{figure}

To verify the physical validity of the estimated ROA under large grid disturbances, EMT simulations are performed. In physical power systems, a large-signal disturbance, such as a temporary grid fault, injects kinetic energy into the dynamics and perturbs the system from its equilibrium to a post-fault initial state $\mathbf{x}_{0}$. Fig.~\ref{fig:roa_9bus} compares the ROAs estimated by different methods and contrasts the time-domain dynamical responses of states perturbed near the estimated boundary. When a severe disturbance perturbs the post-fault state to a location inside the proposed boundary (green star marker), the time-domain simulation confirms successful fault ride-through. The phase trajectory (bottom right) exhibits damped non-linear oscillations that robustly spiral inward to the stable equilibrium point. Conversely, when a slightly more critical fault displaces the initial state just outside the proposed boundary (red dot), physical synchronism is lost. The corresponding phase angle trajectory (top right) undergoes monotonic divergence. This sharp dynamical contrast verifies that the estimated boundary accurately captures the physical transient stability limit.
\begin{table}[t]
\centering
\caption{Quantitative ROA Coverage Comparison for Case I}
\label{tab:roa_smib}
\setlength{\tabcolsep}{12pt}      
\renewcommand{\arraystretch}{1.4} 
\begin{tabular}{lc}
\toprule
Method & ROA Size (\%) \\
\midrule
LQR             & 2.23 \\
Lipschitz       & 13.80 \\
Proposed Method & \textbf{44.63}\\
\bottomrule
\end{tabular}
\end{table}
\begin{table}[t]
\centering
\caption{System Parameters for Case II}
\label{tab:param_caseII}
\setlength{\tabcolsep}{4pt}
\renewcommand{\arraystretch}{1.25}
\begin{tabular}{@{}c cccc ccc@{}}
\toprule
\multicolumn{5}{c}{GFM Parameters} & \multicolumn{3}{c}{Network Parameters} \\
\cmidrule(r){1-5}\cmidrule(l){6-8}
GFM   & $H_i$ & $D_i$ & $E_i$ & $P_i^{\ast}$ & Line     & $R_{ij}$ & $X_{ij}$ \\
Index & (s)   & (p.u.)& (p.u.)& (p.u.)       & $i$--$j$ & (p.u.)   & (p.u.)   \\
\midrule
1 & 8 & 1.6 & 1.0566 & 0.7163 & 1--2 & 0.1210 & 0.6380 \\
2 & 6 & 1.2 & 1.0502 & 1.6300 & 1--3 & 0.1357 & 0.7924 \\
3 & 5 & 1.0 & 1.0170 & 0.8500 & 2--3 & 0.1733 & 0.8852 \\
\midrule
\multicolumn{8}{c}{\footnotesize $M_i=2H_i$, $D_i=0.1M_i$;}\\
\bottomrule
\end{tabular}
\end{table}

For quantitative assessment, the area of each estimated ROA is computed via Monte-Carlo integration utilizing $2\times10^{5}$ samples drawn uniformly from the bounded state-space region $[-1,1]^2$. As reported in Table~\ref{tab:roa_smib}, the classical LQR certificate captures merely 2.23\% of this evaluated state space, while the analytical Lipschitz-certified base ROA covers 13.80\%. In contrast, the proposed method successfully certifies an area accounting for 44.63\% of the domain. This achieves a more than $3.2\times$ enlargement over the Lipschitz analytical certificate at a minimal computational overhead, requiring fewer than 50 time-domain EMT evaluations. In contrast, resolving the same boundary by exhaustive EMT evaluation on a uniform grid at resolution $h=0.05$ would require $40^{2}=1600$ simulations over $[-1,1]^{2}$.

\subsection{Case II: Three-GFM System in the COI Frame}
To evaluate scalability in multi-machine power electronics-dominated grids, the proposed framework is applied to the 3-GFM system depicted in Fig.~\ref{3cases}\,(b). The dynamics are mapped to the COI reference frame. This transformation yields a four-dimensional reduced state vector $\mathbf{x} \in \mathbb{R}^{4}$ comprising relative angles and speed deviations. Complete system parameters are detailed in Table \ref{tab:param_caseII}.

\begin{figure}
\centering
\includegraphics[width=1\linewidth]{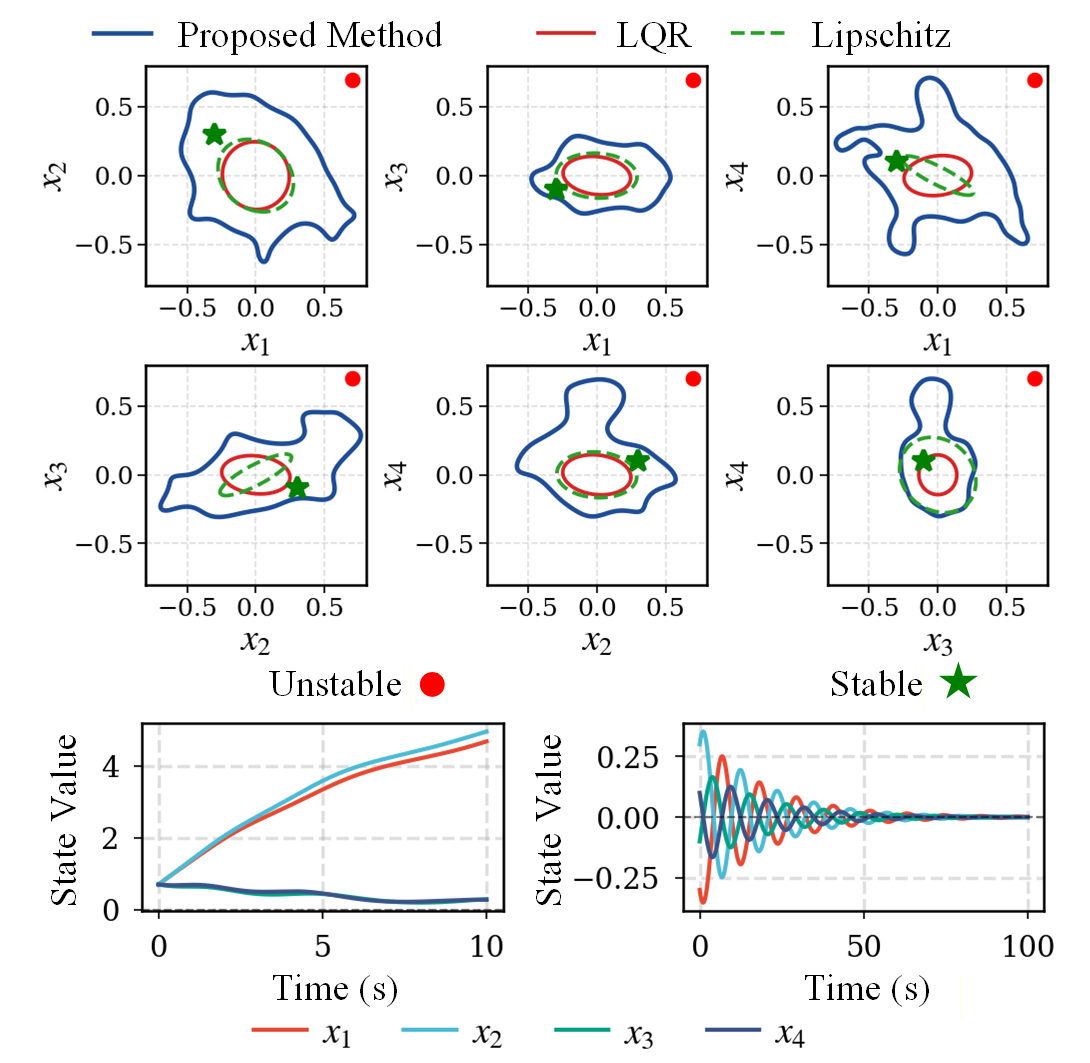}
\caption{ROA comparison and time-domain verification for Case II.}
\label{fig:topo_9bus}
\vspace{-0.75cm}
\end{figure}

Applying the proposed framework yields the expanded four-dimensional ROA, visualized across six two-dimensional coordinate slices in Fig.~\ref{fig:topo_9bus}. Unlike classical LQR and Lipschitz certificates that conservatively restrict the stability boundary to a rigid ellipse around the equilibrium point, the proposed framework flexibly captures the true nonlinear operational margins. Physical time-domain EMT validations corroborate these boundaries: perturbed initial states inside the proposed ROA successfully damp out multi-machine oscillations (bottom right of Fig.~\ref{fig:topo_9bus}), whereas states perturbed outside undergo multi-machine separation and unstable divergence (bottom left of Fig.~\ref{fig:topo_9bus}).

\begin{table}
\centering
\caption{Per-Slice ROA Coverage for Case II}
\label{tab:roa_3mg}
\begin{tabular}{lccc}
\toprule
Slice & LQR (\%) & Lipschitz (\%) & Proposed (\%) \\
\midrule
$(x_1,x_2)$ & 7.45 & 8.83 & 36.63 \\
$(x_1,x_3)$ & 4.21 & 5.95 & 15.63 \\
$(x_1,x_4)$ & 4.41 & 2.39 & 30.39 \\
$(x_2,x_3)$ & 4.26 & 2.64 & 23.77 \\
$(x_2,x_4)$ & 4.47 & 6.00 & 23.96 \\
$(x_3,x_4)$ & 2.41 & 9.36 & 13.81 \\
\midrule
\textbf{Mean} & 4.54 & 5.86 & \textbf{24.03} \\
\midrule
\end{tabular}
\end{table}
\begin{figure*}
\centering
\includegraphics[width=0.95\linewidth]{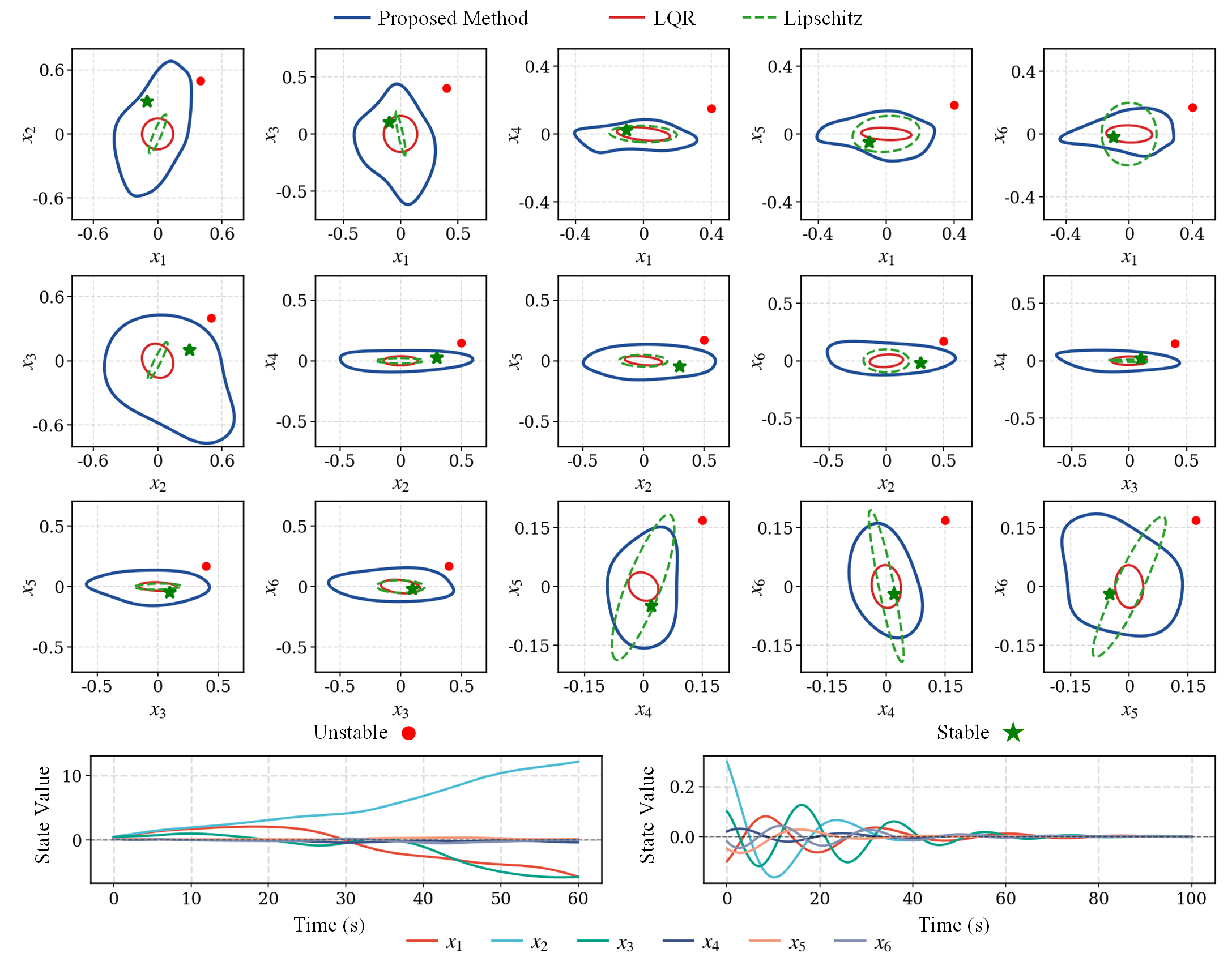}
\caption{ROA comparison and time-domain verification for Case III.}
\label{fig:roa_4mg}
\end{figure*}
\begin{table}[t]
\centering
\caption{System Parameters for Case III}
\label{tab:param_caseIII}
\setlength{\tabcolsep}{4pt}
\renewcommand{\arraystretch}{1.25}
\begin{tabular}{@{}c cccc ccc@{}}
\toprule
\multicolumn{5}{c}{GFM Parameters} & \multicolumn{3}{c}{Network Parameters} \\
\cmidrule(r){1-5}\cmidrule(l){6-8}
GFM   & $H_i$ & $D_i$ & $E_i$ & $P_i^{\ast}$ & Line     & $R_{ij}$ & $X_{ij}$ \\
Index & (s)   & (p.u.)& (p.u.)& (p.u.)       & $i$--$j$ & (p.u.)   & (p.u.)   \\
\midrule
1 & 8 & 1.6 & 1.05 & 0.00 & 1--2 & 1.2030 & 1.1034 \\
2 & 6 & 1.2 & 1.04 & 0.20 & 1--3 & 1.0300 & 0.7400 \\
3 & 5 & 1.0 & 1.03 & 0.15 & 3--4 & 1.5042 & 1.3554 \\
4 & 4 & 0.8 & 1.02 & 0.18 &      &        &        \\
\midrule
\multicolumn{8}{c}{\footnotesize $M_i=2H_i$, $D_i=0.1M_i$; }\\
\bottomrule
\end{tabular}
\end{table}
Quantitative evaluation across each two-dimensional slice is detailed in Table~\ref{tab:roa_3mg}, computed via Monte-Carlo integration over the $[-0.8,0.8]^2$ spatial domain. Across all evaluated coordinate slices, the proposed method achieves an average area coverage of 24.03\%. As visually summarized in Fig.~\ref{fig:topo_4mg}, this achieves a substantial $4.1\times$ enlargement over the analytical Lipschitz baseline and a $5.3\times$ enlargement over the LQR certificate, accomplished efficiently utilizing a budget of only 220 time-domain EMT evaluations. By comparison, exhaustive EMT evaluation of the same four-dimensional domain at identical resolution would demand $32^{4}\approx1.0\times10^{6}$ simulations.


\subsection{Case III: Four-GFM System in the COI Frame}
The final and most challenging case is a radial network of four GFM units with heterogeneous inertia constants, interconnected through three lossy distribution lines, as shown in Fig.~\ref{3cases}\,(c). Following the COI transformation, the system dynamics are represented by a six-dimensional state vector $\mathbf{x}=[\Delta\delta_{1},\Delta\delta_{2},\Delta\delta_{3},\nu_{1},\nu_{2},\nu_{3}]^{\top} \in \mathbb{R}^{6}$, capturing relative rotor angles and frequency deviations. Complete system parameters are detailed in Table \ref{tab:param_caseIII}.

Applying the proposed framework maps the six-dimensional stability boundary across fifteen two-dimensional coordinate slices, as depicted in Fig.~\ref{fig:roa_4mg}. To substantiate these estimated boundaries, EMT simulations are subjected to severe grid faults. Following severe large-signal disturbances, post-fault initial states displaced into the estimated ROA successfully absorb the transient energy and converge back to the stable equilibrium origin via damped multi-machine oscillations (bottom right of Fig.~\ref{fig:roa_4mg}). Conversely, when fault events displace post-fault states outside the estimated boundary, internal rotor angles monotonically diverge from the origin (bottom left of Fig.~\ref{fig:roa_4mg}).

\begin{table}
\centering
\vspace{-0.3cm}
\caption{Per-Slice ROA Coverage for Case III}
\label{tab:roa_4mg}
\begin{tabular}{lccc}
\toprule
Slice & LQR (\%) & Lipschitz (\%) & Proposed (\%) \\
\midrule
$(x_1,x_2)$ & 2.56 & 0.91 & 23.88 \\
$(x_1,x_3)$ & 3.19 & 0.62 & 21.22 \\
$(x_1,x_4)$ & 1.84 & 2.94 & 9.88 \\
$(x_1,x_5)$ & 1.67 & 6.59 & 13.76 \\
$(x_1,x_6)$ & 2.15 & 9.40 & 12.09 \\
$(x_2,x_3)$ & 2.83 & 0.77 & 41.18 \\
$(x_2,x_4)$ & 0.86 & 0.71 & 8.01 \\
$(x_2,x_5)$ & 0.89 & 1.56 & 12.75 \\
$(x_2,x_6)$ & 1.15 & 2.83 & 11.59 \\
$(x_3,x_4)$ & 0.84 & 0.33 & 6.87 \\
$(x_3,x_5)$ & 0.90 & 0.79 & 11.15 \\
$(x_3,x_6)$ & 1.41 & 1.66 & 11.31 \\
$(x_4,x_5)$ & 2.22 & 13.96 & 23.28 \\
$(x_4,x_6)$ & 3.33 & 6.32 & 21.18 \\
$(x_5,x_6)$ & 3.16 & 11.21 & 38.77 \\
\midrule
\textbf{Mean} & 1.93 & 4.04 & \textbf{17.80} \\
\midrule
\end{tabular}
\vspace{-0.6cm}
\end{table}
\begin{figure}[!t]
\vspace{-0.8cm}
\centering
\includegraphics[width=0.95\linewidth]{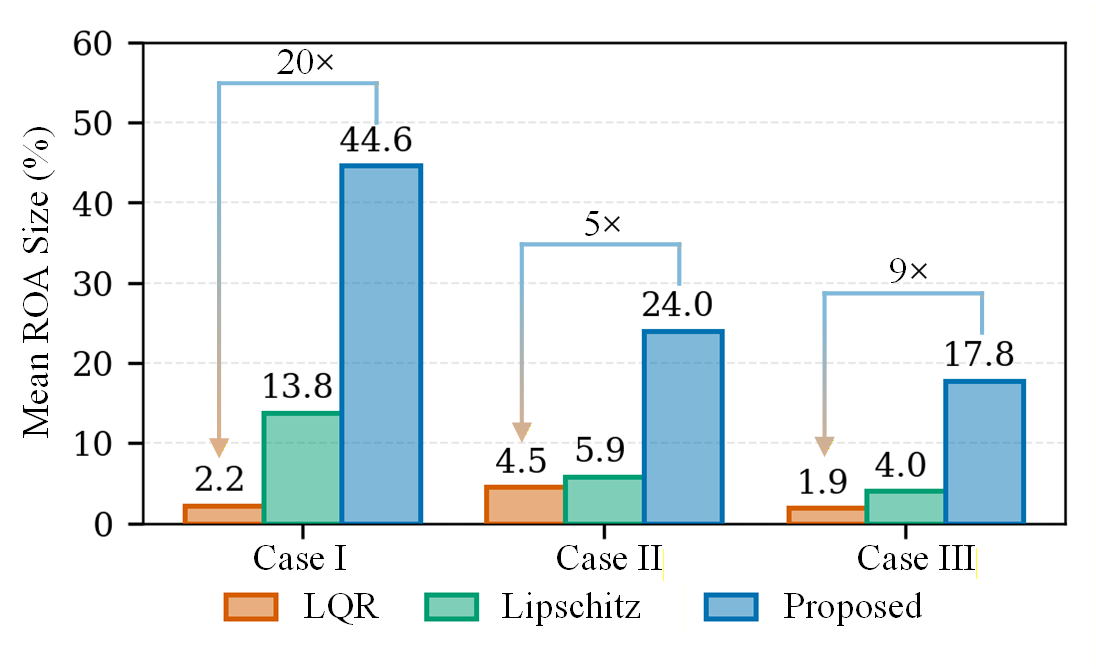}
\caption{Quantitative comparison of ROA sizes across three case studies.}
\label{fig:topo_4mg}
\end{figure}

Quantitative evaluation across all fifteen slices is reported in Table~\ref{tab:roa_4mg}, computed via Monte-Carlo integration utilizing $1\times10^{5}$ uniform samples per slice window. The proposed method achieves a mean area coverage of 17.80\%, representing a $4.4\times$ enlargement over the analytical Lipschitz base region and a substantial $9.2\times$ expansion over the classical LQR certificate. As summarized in Fig.~\ref{fig:topo_4mg}, the relative expansion capacity of the proposed algorithm remains highly consistent as the state-space dimension increases from two to six. By actively anchoring EMT queries directly on the uncertainty frontier rather than searching the entire state space, the proposed framework successfully certifies this high-dimensional boundary using a compact budget of only 220 time-domain EMT evaluations. Exhaustive EMT evaluation of this six-dimensional domain at the same resolution would instead require $32^{6}\approx1.1\times10^{9}$ simulations.

\section{Conclusion}
This paper has presented a three-stage probabilistic active learning framework that substantially reduces the conservativeness of ROA estimation for grid-interactive inverter systems. The framework is anchored on a deterministically guaranteed base ROA, synthesized as a neural Lyapunov function and certified through Lipschitz-based algebraic verification. Because any further enlargement of a neural Lyapunov-based estimate is obstructed by gradient distortion and pseudo-counterexamples in the OOD region, the proposed method abandons the decrease condition outside the certified core and directly evaluates the physical stability boundary via a GP regression model. 

Case studies spanning two- to six-dimensional benchmarks validate the scalability and accuracy of the framework. On the single-GFM system, the estimated region certifies $44.63\%$ of the evaluated domain, achieving a $3.2$-fold enlargement over the $13.80\%$ Lipschitz baseline and significantly outperforming the $2.23\%$ LQR certificate. On the three-GFM system and the four-GFM system, the methodology achieves mean volumetric expansions of $4.1$ and $4.4$ times over analytical Lipschitz certificates, and $5.3$ and $9.2$ times over classical LQR baselines. By actively anchoring exploration directly on boundary ridges, the proposed method captures the less conservative ROA at the cost of up to $220$ time-domain simulation queries per system, demonstrating high computational efficiency. Future work will extend this proposed framework toward black-box GFM inverters with inaccessible internal dynamic models, as well as complex hybrid power systems comprising heterogeneous generation units.

\vspace{-0.4cm}
\appendices
\section{COI Reference Frame Transformation}
\label{app:coi}
The center-of-inertia (COI) of the $N$-unit system is defined by the inertia-weighted angle and frequency
\begin{equation}
\theta_{\mathrm{COI}} = \frac{1}{M_{T}}\sum_{i=1}^{N} M_{i}\,\delta_{i}, \qquad
\omega_{\mathrm{COI}} = \frac{1}{M_{T}}\sum_{i=1}^{N} M_{i}\,\omega_{i},
\label{eq:coi_def}
\end{equation}
with total inertia $M_{T}=\sum_{i=1}^{N} M_{i}$. Each unit is referred to the COI through $\tilde{\delta}_{i}=\delta_{i}-\theta_{\mathrm{COI}}$ and $\nu_{i}=\omega_{i}-\omega_{\mathrm{COI}}$, and its angle deviation from the post-fault SEP is $\Delta\delta_{i}=\tilde{\delta}_{i}-\tilde{\delta}_{i}^{\ast}$. By construction, the COI coordinates satisfy the two constraints
\begin{equation}
\sum_{i=1}^{N} M_{i}\,\Delta\delta_{i} = 0, \qquad \sum_{i=1}^{N} M_{i}\,\nu_{i} = 0,
\label{eq:coi_constraint}
\end{equation}
so the state of any one unit, say unit $N$, is recovered from the remaining $N-1$ units:
\begin{equation}
\Delta\delta_{N} = -\frac{1}{M_{N}}\sum_{i=1}^{N-1} M_{i}\,\Delta\delta_{i}, \qquad
\nu_{N} = -\frac{1}{M_{N}}\sum_{i=1}^{N-1} M_{i}\,\nu_{i}.
\label{eq:coi_recover}
\end{equation}
Subtracting the COI acceleration from each swing equation \eqref{eq:swing2} gives the reduced COI dynamics
\begin{align}
\dot{\Delta\delta}_{i} &= \nu_{i}, \\
M_{i}\,\dot{\nu}_{i} &= \bigl(P_{i}^{\ast}-P_{i}^{\mathrm{e}}-D_{i}\nu_{i}\bigr)
- \frac{M_{i}}{M_{T}}\sum_{j=1}^{N}\bigl(P_{j}^{\ast}-P_{i}^{\mathrm{e}}-D_{j}\nu_{j}\bigr),
\label{eq:coi_dyn}
\end{align}
for $i=1,\dots,N-1$. The subtracted term is the COI acceleration, which guarantees $\sum_{i=1}^{N} M_{i}\dot{\nu}_{i}=0$ and thereby removes the redundant common mode. Together with \eqref{eq:coi_recover}, this yields the autonomous reduced-order model in \eqref{eq:state_vector} on which the proposed framework operates.


 
%
\bibliographystyle{IEEEtran} 
\bibliography{ref}

\end{document}